\documentclass[aps,prl,twocolumn,superscriptaddress,nofootinbib]{revtex4-2}
\usepackage{graphicx}
\usepackage{amsmath,amssymb}
\usepackage{bm}
\usepackage{hyperref}

\begin{document}

\title{Inelastic Dark Photon Dark Matter for\\ the LUX-ZEPLIN High-Recoil Event and the Galactic Halo Gamma-Ray Excess}

\author{Kimiko Yamashita}
\email{kimiko.yamashita.nd93@vc.ibaraki.ac.jp}
\affiliation{Department of Physics, Ibaraki University, Mito 310-8512, Japan}

\date{\today}

\begin{abstract}
We show that a dark-photon dark matter framework can simultaneously
account for two recently reported anomalies: the halo-like Galactic
gamma-ray excess identified by Totani in 15 years of Fermi-LAT data, and
the $E_{\rm nr}=248\pm23\,({\rm stat})\pm23\,({\rm sys})$~keV nuclear-recoil
event of interest reported by the LUX-ZEPLIN Collaboration.
The dark matter is a $U(1)_X$ dark photon $X$ with mass $m_X=420$~GeV,
stabilized by a dark parity that forbids kinetic mixing with the Standard
Model photon. A light scalar mediator $\phi$, required to explain
the Totani excess via Sommerfeld enhancement, is joined by a second
light scalar $\phi'$ that couples a
nearly degenerate heavier vector partner $V$ to $X$. The resulting
inelastic transition $X+N\to V+N$, with mass splitting
$\delta\simeq316$~keV close to the kinematic endpoint for $m_X=420$~GeV,
naturally reproduces the localized LZ event once the detector energy
resolution is included, while leaving the relic abundance and the halo
phenomenology of the original Totani-motivated benchmark essentially
unchanged.
\end{abstract}

\maketitle

\textit{Introduction.}---%
Recently, a high-recoil-energy analysis by the LUX-ZEPLIN (LZ) experiment
has reportedly identified an event over the expected background in an
extended nuclear-recoil energy window, $E_{\rm nr}=248\pm23\,({\rm stat})
\pm23\,({\rm sys})$~keV, corresponding to a local significance of $3.4\sigma$
and a global significance of $2.6\sigma$~\cite{LZ2026}. A localized
feature at such high recoil energy is not
characteristic of the monotonically falling spectrum expected from
conventional elastic weakly interacting massive particle (WIMP)-nucleus
scattering, but can instead be selected by inelastic scattering through its
threshold kinematics. Several particle-physics interpretations of this
event have already been proposed, invoking endothermic dark
matter~\cite{Su2026}, Higgsino dark matter~\cite{Freese2026,Wu2026}, and
fermionic dark matter absorption~\cite{Lou2026}.

Independently, Totani analyzed 15 years of Fermi Large Area Telescope
data and reported a halo-like excess in the Galactic diffuse gamma-ray
emission, well fitted by dark matter annihilation into $W^+W^-$ with
$m_X=420$~GeV, but requiring a present-day cross section
$\langle\sigma v\rangle_{\rm halo}\sim10^{-24}\,{\rm cm}^3\,{\rm s}^{-1}$
that exceeds the canonical thermal-relic value by about two orders of
magnitude and the dwarf-spheroidal limit,
$\langle\sigma v\rangle_{\rm dwarf}\lesssim2\times10^{-25}\,{\rm
cm}^3\,{\rm s}^{-1}$, by a factor of several~\cite{Totani2025}. This
tension can be resolved by exploiting the hierarchy of dark matter
velocities between freeze-out, the Galactic halo, and dwarf spheroidal
galaxies~\cite{Murayama2025,NomuraTotani2026}: a $p$-wave-suppressed
cross section fixes the correct relic abundance while evading dwarf and
cosmic microwave background bounds, and a Sommerfeld enhancement that
grows toward low velocity and saturates in the halo boosts the
present-day rate, as realized in
Refs.~\cite{Jho2025,Yoshimatsu2026,Yamashita2026}. In previous
work~\cite{Yamashita2026} we implemented this mechanism with a
parity-violating Higgs-portal dark
photon~\cite{Yamashita2024,DeFelice2025} and a light scalar mediator
$\phi$, accounting for the Totani excess at $m_X=420$~GeV consistently
with dwarf-spheroidal and cosmic-microwave-background constraints.

In this Letter, we show that the same framework, extended by a
second, nearly degenerate real vector $V$ and an accompanying light
scalar $\phi'$, naturally accommodates the LZ
event as well. The mass splitting $\delta\equiv m_V-m_X$ required to place
the kinematic endpoint of the inelastic recoil spectrum at the observed
energy is close to the maximal splitting kinematically accessible for
$m_X=420$~GeV, so that the same dark matter mass that explains the Totani
excess also naturally explains the LZ event once the finite detector energy
resolution is taken into account. Because the new interactions are
comparatively weak, they leave the coannihilation-corrected relic abundance
and the Sommerfeld-enhanced halo annihilation rate close to their values in
Ref.~\cite{Yamashita2026}.

\textit{Model.}---%
We consider a dark photon $X_\mu$, the gauge boson of a $U(1)_X$ symmetry
under which all Standard Model (SM) fields are neutral. A dark parity,
under which $X_\mu$ is odd, forbids kinetic mixing with SM hypercharge, so
that $X$ is absolutely stable. The leading portal interactions with the SM
are the dimension-six operators $\mathcal{O}=(H^\dagger
H)X_{\mu\nu}X^{\mu\nu}$ and $\tilde{\mathcal
O}=(H^\dagger H)X_{\mu\nu}\tilde X^{\mu\nu}$~\cite{Yamashita2024}, with
$X_{\mu\nu}=\partial_\mu X_\nu-\partial_\nu X_\mu$ and
$\tilde X^{\mu\nu}=\frac12\epsilon^{\mu\nu\rho\sigma}X_{\rho\sigma}$. We
focus on the parity-odd operator,
\begin{equation}
\mathcal{L}\supset\frac{\tilde C}{\Lambda^2}\tilde{\mathcal O}
=\frac{\tilde C}{\Lambda^2}(H^\dagger H)\,X_{\mu\nu}\tilde X^{\mu\nu},
\label{eq:portal}
\end{equation}
since it simultaneously (i) yields the $p$-wave-suppressed annihilation
required to reconcile the Totani excess with the thermal relic abundance
and dwarf-spheroidal and cosmic microwave background constraints, and
(ii) renders the tree-level direct-detection amplitude of $X$ on nuclei
zero at vanishing momentum transfer, evading the stringent limits from
conventional elastic WIMP searches that would otherwise apply to the
parity-even operator $\mathcal{O}$~\cite{Yamashita2024}. A light CP-even
scalar $\phi$
coupled to the dark-photon mass operator,
$\mathcal{L}\supset\frac12\mu\,\phi\,X_\mu X^\mu$, mediates an attractive
Yukawa force between $X$ particles, producing the Sommerfeld-enhanced
halo annihilation rate of Table~\ref{tab:benchmark} without affecting
freeze-out~\cite{Yamashita2026}.

\begin{table}[t]
\centering
\caption{Benchmark parameters and representative phenomenological
quantities used in this work, taken from Ref.~\cite{Yamashita2026}
(v2).}
\label{tab:benchmark}
\begin{tabular}{lc}
\hline\hline
Quantity & Value \\
\hline
DM mass $m_X$ & $420~\mathrm{GeV}$ \\
Mediator mass $m_\phi$ & $400~\mathrm{MeV}$ \\
Trilinear coupling $\mu$ & $1.3~\mathrm{TeV}$ \\
Yukawa-potential strength $\alpha_{\rm eff}(m_X,\mu)$ & $0.2$ \\
Cutoff scale $\Lambda$ & $1~\mathrm{TeV}$ \\
\hline
Freeze-out velocity $v_{\rm fo}$ & $0.3$ \\
Halo velocity $v_{\rm halo}$ & $10^{-3}$ \\
Dwarf velocity $v_{\rm dwarf}$ & $10^{-4}$ \\
\hline
$S_1(v_{\rm fo})$ & $6.1$ \\
$S_1(v_{\rm halo})$ & $1.3\times10^{7}$ \\
$S_1(v_{\rm dwarf})$ & $3.1\times10^{7}$ \\
\hline
$\langle\sigma v\rangle_{\rm fo}^{W^+W^-}$ & $1.9\times10^{-26}~\mathrm{cm}^3\,\mathrm{s}^{-1}$ \\
$\langle\sigma v\rangle_{\rm halo}$ & $2.7\times10^{-24}~\mathrm{cm}^3\,\mathrm{s}^{-1}$ \\
$\langle\sigma v\rangle_{\rm dwarf}$ & $6.5\times10^{-26}~\mathrm{cm}^3\,\mathrm{s}^{-1}$ \\
\hline\hline
\end{tabular}
\end{table}

To accommodate the LZ event, we extend the dark sector by a second real
vector $V_\mu$, nearly degenerate with $X$ and likewise odd under the
dark parity, with mass splitting
$\delta\equiv m_V-m_X\ll m_X$ treated as a free parameter. The full set
of dimension-six operators allowed by the symmetries of the model
includes, in addition to $\mathcal O,\tilde{\mathcal O}$ applied to $V$,
the off-diagonal terms $(H^\dagger H)X_{\mu\nu}V^{\mu\nu}$,
$(H^\dagger H)X_{\mu\nu}\tilde V^{\mu\nu}$, and
$B_\mu^{\ \nu}X_\nu^{\ \alpha}V_\alpha^{\ \mu}$,
$\tilde B_\mu^{\ \nu}X_\nu^{\ \alpha}V_\alpha^{\ \mu}$~\cite{Aebischer2022},
where $B^{\mu\nu}$ is the $U(1)_Y$ field strength. Mediating the
inelastic transition through the same scalar $\phi$ that drives
Sommerfeld enhancement would also induce elastic $X+N\to X+N$
scattering through the diagonal vertex $\mu\,\phi\,X_\mu X^\mu$ once
$\phi$-$h$ mixing is switched on, with $\mu=1.3$~TeV vastly exceeding
$\mu_{XV}$ and no inelastic threshold to suppress it; this would
overshoot current elastic spin-independent limits by many orders of
magnitude. We therefore introduce a second light scalar $\phi'$,
degenerate with $\phi$ in mass but coupling only off-diagonally,
$\mu_{XV}\phi' X_\mu V^\mu$, and mixing with the Higgs through
$\mu_{\phi'h}\phi' H^\dagger H$; $\phi$ itself does not mix with $h$.
Since $X$ and $V$ are both odd under the dark parity introduced
above, $X_\mu V^\mu$ is parity-even, just like $X_\mu X^\mu$ and
$V_\mu V^\mu$; dark parity alone therefore cannot distinguish these
three bilinears from one another, and a discrete symmetry beyond it is
needed to select $\phi$ to couple only to the diagonal combination and
$\phi'$ only to $X_\mu V^\mu$, and to forbid $\phi H^\dagger H$. Of
these requirements, two
are phenomenologically essential -- the absence of $\phi' X_\mu X^\mu$
and of $\phi H^\dagger H$, since either would reopen the elastic
direct-detection danger for $X$ -- while the absence of $\phi' V_\mu
V^\mu$ and
$\phi X_\mu V^\mu$ is not: they involve only the short-lived $V$ or
the (otherwise) Higgs-decoupled $\phi$. A full symmetry construction
consistent
with the original portal operators of Eq.~\eqref{eq:portal} is left
for future work. Because
the two-body channel $V\to X\phi'$ is kinematically closed
($\delta<m_{\phi'}$), and
decay through the Higgs-portal cross terms is suppressed by the far
off-shell Higgs propagator ($m_h\gg\delta$) while decay via an
off-shell $\phi'^*$ is both severely phase-space suppressed and further
suppressed by the small $\phi'$-$h$ mixing angle and the loop level of
the inherited $h\to\gamma\gamma$ coupling through which it alone reaches
the diphoton final state, neither is fast enough to complete before big
bang nucleosynthesis (BBN); we retain only the hypercharge operator
$B_\mu^{\ \nu}X_\nu^{\ \alpha}V_\alpha^{\ \mu}$, since the parity-odd
partner $\tilde B_\mu^{\ \nu}X_\nu^{\ \alpha}V_\alpha^{\ \mu}$ requires
the Levi-Civita contraction to route the large $\mathcal{O}(m_X)$
energy of the at-rest $V$ through the small spatial recoil of $X$,
suppressing its decay rate by a further $(\delta/m_X)^2$,\footnote{The
totally antisymmetric $\epsilon$ tensor forbids both the $V$ and $X$
legs from carrying their large time components simultaneously; since
$V$ is at rest, this forces the small $\mathcal{O}(\delta)$ spatial
momentum of $X$ into the leg that would otherwise carry $m_X$, unlike
the non-tilde operator, where no such cancellation occurs.} which
contains $B_{\mu\nu}\supset\cos\theta_W F_{\mu\nu}$ after electroweak
symmetry breaking and induces a tree-level two-body decay $V\to X\gamma$
with $\Gamma\propto\delta^3m_X^2/\Lambda^4$; for an $\mathcal{O}(1)$
coefficient this gives $\tau_V\sim3\,\mu$s, safely before BBN. Rapid
depletion of $V$ is in fact doubly required: if $V$ survived to the
present epoch, the exothermic process $V+N\to X+N$ would proceed
unsuppressed even at $v_{\rm min}=0$, unlike the
velocity-tail-suppressed endothermic signal discussed below, and would
overwhelm the observed rate. The
full interaction Lagrangian governing the Totani excess, the LZ
signal, and the relic abundance reads
\begin{align}
\mathcal{L}_{\rm int}=&\,\frac{\tilde C}{\Lambda^2}(H^\dagger H)\,
X_{\mu\nu}\tilde X^{\mu\nu}
+\frac12\mu\,\phi\,(X_\mu X^\mu+V_\mu V^\mu)\notag\\
&+\mu_{XV}\,\phi'\,X_\mu V^\mu+\mu_{\phi'h}\,\phi'\,H^\dagger H,
\label{eq:3-point}
\end{align}
where $\mu_{XV}$ provides a tree-level, momentum-independent
$X$-$V$-$\phi'$ vertex, and the dimension-three mixing $\mu_{\phi'h}$
connects $\phi'$ to SM fermions through a small $\phi'$-$h$ mixing
angle $\sin\theta'\simeq\mu_{\phi'h}v/m_h^2$. Since $m_{\phi'}=400$~MeV
exceeds $m_K-m_\pi$, the decay $K^+\to\pi^+\phi'$ is kinematically
closed; we work in the regime $\sin\theta'\lesssim$ a few$\times10^{-2}$,
consistent with existing constraints on light Higgs-mixed
scalars~\cite{HiggsMixingConstraints}. The transition $X+N\to V+N$,
mediated by $\phi'$ exchange through Eq.~\eqref{eq:3-point}, provides
the inelastic direct-detection signal discussed below, while leaving
the diagonal couplings that control the Totani excess and the relic
abundance unchanged.

\textit{Relic Abundance and Coannihilation.}---%
Because $\delta\sim\mathcal{O}(100)$~keV is many orders of magnitude
below the freeze-out temperature $T_{\rm fo}\sim m_X/20\sim21$~GeV, the
Boltzmann suppression factor $e^{-\delta/T_{\rm fo}}\approx1$: $X$ and $V$
remain in chemical equilibrium throughout freeze-out and must be treated
as a single coannihilating system with total effective degrees of
freedom $g_{\rm eff}=g_X+g_V=6$, twice that of the single-vector
calculation of Ref.~\cite{Yamashita2026}. The dominant annihilation
channels $XX,VV\to W^+W^-$ proceed through the
Higgs-portal of Eq.~\eqref{eq:portal}, applied symmetrically to $X$
and $V$, and remain $p$-wave suppressed. The new off-diagonal channel
$XV\to\phi'^*\to f\bar f$, mediated by the $\mu_{XV}$ vertex followed by
$\phi'$-$h$ mixing, proceeds through a $\phi'$ propagator deeply off
resonance ($\sqrt s\simeq2m_X\gg m_{\phi'}$) and is suppressed by both
$1/s$ and $\sin^2\theta'$; we choose $\mu_{XV}$ such that its contribution
to $\langle\sigma v\rangle_{\rm fo}$ remains below a few percent of the
$W^+W^-$ channel. With $X$ and $V$ nearly degenerate,
$n_X^{\rm eq}\approx n_V^{\rm eq}\approx n^{\rm eq}/2$, so the
thermally averaged annihilation cross section is diluted to
$\langle\sigma_{\rm eff}v\rangle\approx\frac12\langle\sigma_{XX}v\rangle$;
reproducing $\Omega_{\rm DM}h^2=0.12$ then requires
$\tilde C\to\sqrt{2}\,\tilde C\simeq1.4$, given $\tilde C=1$ in
Ref.~\cite{Yamashita2026}, still $\mathcal{O}(1)$.
Because $V$ has decayed away by the time structure
forms, the present-day Galactic halo is populated by $X$ alone, and the
Sommerfeld-enhanced annihilation rates of Table~\ref{tab:benchmark} are
unaffected.

\begin{figure}[t]
\centering
\includegraphics[width=0.95\linewidth]{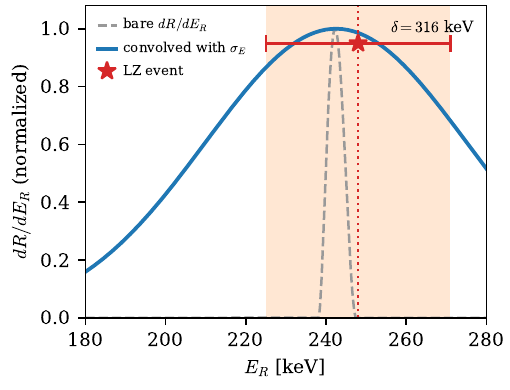}
\caption{Differential inelastic recoil spectrum at zero momentum
transfer, $F_{\rm DM}(q^2\to0)=1$, for $\delta=316$~keV, before (dashed)
and after (solid) convolution with the LZ energy resolution, normalized
to unit maximum. The shaded band and star show the observed event,
$E_{\rm nr}=248\pm32.5$~keV (statistical and systematic uncertainties
combined in quadrature).}
\label{fig:dRdER}
\end{figure}

\textit{Inelastic Scattering at LZ.}---%
The transition $X+N\to V+N$ proceeds through $t$-channel $\phi'$
exchange via the $\mu_{XV}$ vertex of Eq.~\eqref{eq:3-point}, followed
by $\phi'$-$h$ mixing and the standard spin-independent Higgs-nucleon
coupling $f_N\approx0.3$. The minimum dark matter speed required to
produce a nuclear recoil $E_R$ is~\cite{TuckerSmith2001}
\begin{equation}
v_{\rm min}=\left|\frac{m_A E_R}{\mu_{XA}}+\delta\right|
\frac{1}{\sqrt{2m_AE_R}},
\label{eq:vmin}
\end{equation}
with $\mu_{XA}$ the $X$-nucleus reduced mass; the endothermic ($+\delta$)
sign forbids recoils below a threshold that grows with $\delta$, while
$v_{\rm min}$ reaches the Galactic escape velocity
$v_{\rm max}\equiv v_{\rm esc}+v_e$ at a maximal recoil energy,
producing a pinch in the differential rate. This pinch disappears once
$\delta$ exceeds the kinematic ceiling
$\delta_{\max}=\mu_{XA}v_{\rm max}^2/2$, which for $m_X=420$~GeV
evaluates to $\delta_{\max}=316$~keV for the heaviest natural xenon
isotope, $^{136}$Xe; our benchmark $\delta\simeq316$~keV therefore
saturates this bound. At the standard zero-momentum-transfer reference
point, $F_{\rm DM}(q^2\to0)=1$, where
$F_{\rm DM}(q)\equiv|\mathcal{M}(q)|^2/|\mathcal{M}(q=0)|^2$ is the
mediator form factor, the differential rate is~\cite{LewinSmith1996}
\begin{equation}
\frac{dR}{dE_R}\propto\sum_{A}\xi_A\,A^2F_A^2(q)\,
\eta\big(v_{\rm min}(E_R;\delta)\big),
\end{equation}
where the sum runs over the naturally occurring xenon isotopes with
abundance $\xi_A$, $q=\sqrt{2m_AE_R}$, $F_A(q)$ is the Helm form
factor, and $\eta(v_{\rm min})=\langle1/v\rangle_{v>v_{\rm min}}$ is the
mean inverse speed for a truncated Maxwell-Boltzmann halo
($v_0=220$~km/s, $v_{\rm esc}=544$~km/s, $v_e=220$~km/s), also evaluated
using the closed-form expressions of Ref.~\cite{LewinSmith1996}. We find
that this pinch occurs at $E_R\simeq242$~keV. Restoring the
light-$\phi'$ form factor, $F_{\rm DM}(q)=[m_{\phi'}^2/(q^2+m_{\phi'}^2)]^2$,
changes
$F_{\rm DM}^2$ by only $15\%$ across $E_R\in[200,270]$~keV -- far
milder than the exponential falloff of $\eta(v_{\rm min})$ that
produces the pinch -- and leaves the smeared peak position and the
$E_R=248$~keV rate unchanged to three significant figures. We convolve
the bare spectrum with a Gaussian detector-resolution kernel,
\begin{equation}
\left(\frac{dR}{dE_R}\right)_{\rm smeared}\!\!=\!\int\! dE_R'\,
\frac{dR}{dE_R'}\,\frac{e^{-(E_R-E_R')^2/2\sigma_E^2}}{\sqrt{2\pi}\,\sigma_E},
\end{equation}
with $\sigma_E=32.53$~keV combining the quoted statistical and
systematic uncertainties in quadrature; the resulting spectrum peaks at
$E_R\simeq248$~keV with negligible loss relative to its maximum
(Fig.~\ref{fig:dRdER}), matching the observed event once $\delta$ is
fixed, without further tuning.

We fix $\mu_{XV}$ and $\mu_{\phi'h}$ — the latter setting both the
$\phi'$-nucleon coupling $\propto\sin\theta'\,f_N m_N/v$ that controls
the signal strength and the mixing angle itself — to reproduce one
event in the $225$--$271$~keV window for the $2.84$~tonne-year LZ
exposure, subject to the coannihilation bound above and
$\sin\theta'\lesssim$ a few$\times10^{-2}$ (to avoid other
constraints). Table~\ref{tab:LZbenchmark}
summarizes a representative benchmark point; a dedicated
profile-likelihood analysis of the local and global significance is
left for future work.

\begin{table}[t]
\centering
\caption{Benchmark parameters for the LZ signal.}
\label{tab:LZbenchmark}
\begin{tabular}{lc}
\hline\hline
Quantity & Value \\
\hline
DM mass $m_X$ & $420~\mathrm{GeV}$ \\
Mediator mass $m_{\phi'}$ & $400~\mathrm{MeV}$ \\
Mass splitting $\delta$ & $316~\mathrm{keV}$ \\
$X$-$V$-$\phi'$ coupling $\mu_{XV}$ & $2\times10^{2}~\mathrm{GeV}$ \\
$\phi'$-$h$ mixing coefficient $\mu_{\phi'h}$ & $1~\mathrm{GeV}$ \\
Mixing angle $\sin\theta'=\mu_{\phi'h}v/m_h^2$ & $1.6\times10^{-2}$ \\
\hline
$V$ lifetime $\tau_V$ & $3~\mu\mathrm{s}$ \\
$N_{\rm events}$ ($225$--$271$~keV) & $\sim1$ \\
\hline\hline
\end{tabular}
\end{table}

\textit{Discussion.}---%
A dark-photon dark matter framework, built around the parity-odd
Higgs-portal operator that explains the Totani excess through $p$-wave
freeze-out and Sommerfeld enhancement, thus also accounts for the LZ
high-recoil event once extended by a nearly degenerate vector
partner $V$ and an accompanying light scalar $\phi'$. The same mass,
$m_X=420$~GeV, that fits the Totani
spectrum places the kinematic endpoint of the inelastic recoil spectrum
within the LZ energy resolution of the observed event, for $\delta$
close to its kinematically maximal value, without retuning $m_X$; the
coannihilation-corrected relic abundance and present-day annihilation
rate remain close to their values in the single-vector benchmark of
Ref.~\cite{Yamashita2026}. A fully systematic classification of the
two-vector operator basis, potentially realized through a complex
Stueckelberg dark photon~\cite{Bertuzzo2024}, and a discrete-symmetry
construction protecting the $\phi$-$\phi'$ sector, including the
absence of $\phi H^\dagger H$, are
left for future work.
Future LZ exposure, and
the planned CRESST upgrade~\cite{Angloher2026} discussed in
Ref.~\cite{Su2026} for target nuclei heavier than xenon,
will test whether the event of interest persists. Independently, the
Sommerfeld-enhanced $W^+W^-$ continuum responsible for the Totani
excess should be directly testable by the upcoming Cherenkov Telescope
Array Observatory (CTAO)~\cite{Pierre2014}.

\begin{acknowledgments}
This work was supported in part by JSPS KAKENHI Grant Number
JP24K17040.
\end{acknowledgments}

\textit{Declaration of generative AI and AI-assisted technologies in
the manuscript preparation process.}---%
During the preparation of this work, the author used Claude
(Anthropic) to assist with numerical cross-checks of analytic
derivations, literature searches, and drafting and editing of the
manuscript. After using this tool, the author reviewed, verified, and
edited all content as needed and takes full responsibility for the
content of the published article.

\bibliographystyle{apsrev4-2}

\end{document}